\documentclass[twocolumn,floatfix,showpacs,prb,10pt,superscriptaddress]{revtex4-1}
\usepackage{graphicx}
\usepackage{amsmath}
\usepackage{amssymb}
\usepackage{bm}
\usepackage{hyperref}
\usepackage{multirow}
\usepackage{color}
\usepackage{amsbsy}
\usepackage{ulem}
\usepackage[utf8]{inputenc}

\newcommand{\pd}{{\phantom{\dag}}}

\begin{document}

\title{Anomalous Floquet topological crystalline insulators}

\author{Konstantinos Ladovrechis}
\email{konstantinos.ladovrechis@tu-dresden.de}
\affiliation{Institute of Theoretical Solid State Physics, IFW Dresden, 01069 Dresden, Germany}
\affiliation{Institute of Theoretical Physics, Technische Universität Dresden, 01062 Dresden, Germany}

\author{Ion Cosma Fulga}
\affiliation{Institute of Theoretical Solid State Physics, IFW Dresden, 01069 Dresden, Germany}

\date{\today}
\begin{abstract}

Periodically driven systems can host so called anomalous topological phases, in which protected boundary states coexist with topologically trivial Floquet bulk bands. We introduce an anomalous version of reflection symmetry protected topological crystalline insulators, obtained as a stack of weakly-coupled two-dimensional layers. The system has tunable and robust surface Dirac cones even though the mirror Chern numbers of the Floquet bulk bands vanish. The number of surface Dirac cones is given by a new topological invariant determined from the scattering matrix of the system. Further, we find that due to particle-hole symmetry, the positions of Dirac cones in the surface Brillouin zone are controlled by an additional invariant, counting the parity of modes present at high symmetry points.

\end{abstract}
\maketitle

\section{Introduction}
\label{sec:intro}

Topological insulators (TI) are characterized by the presence of a gapped bulk and gapless boundary states.\cite{Hasan2010, Qi2011} The connection between the two can be expressed using bulk-boundary correspondence, which relates the number of protected boundary modes to a topological invariant associated to the bulk eigenstates. First established in connection to the quantum Hall effect,\cite{Thouless1982} bulk-boundary correspondence has since then been extended to a large class of so called symmetry protected topological phases. In the presence of time-reversal, particle-hole, or chiral symmetries, the resulting phases are dubbed strong topological insulators,\cite{Moore2007, Schnyder2008, Roy2009, Kitaev2009} while enforcing symmetries of the lattice, such as translation, rotation, or mirror symmetries, leads to weak topological insulators (WTI)\cite{Fu2007, Slager2012, Jadaun2013, Chiu2013, Zhang2013, Morimoto2013, Seroussi2014, Benalcazar2014, Diez2015} and topological crystalline insulators (TCI).\cite{Teo2008, Fu2011, Hsieh2012}

In parallel with their theoretical prediction, a large variety of topological insulating phases of matter were realized experimentally. Examples include the two- and three-dimensional (2D, 3D) topological insulators protected by time-reversal symmetry,\cite{Konig2007, Hsieh2008, Chen2009, Kuroda2010} 3D reflection symmetry protected TCIs such as SnTe,\cite{Tanaka2012, Dziawa2012, Xu2012} and translation symmetry protected WTIs.\cite{Rasche2013, Pauly2015} Finding new materials or new ways of generating topologically nontrivial behavior remains an area of intense activity in condensed matter physics.\cite{Bradlyn2017}

One promising avenue towards obtaining topological phases of matter is periodic driving. The latter allows to bypass constraints imposed by the chemistry of the material and by the fabrication process, potentially enabling a larger set of topological phases to be realized in the same experimental setup, by adjusting the properties of an externally applied, periodic driving field.\cite{Kitagawa2010, Rudner2013, Quelle2017, Jiang2011, Kundu2013, Reynoso2013, Thakurathi2013, Lababidi2014, Ho2014, Carpentier2015, Klinovaja2016, Leykam2016, Zhou2016, Bomantara2016, Wang2016, Saha2017, Roman-Taboada2017} The resulting phases were dubbed Floquet topological insulators, and have been demonstrated both in 1D and 2D systems, for instance using arrays of coupled photonic waveguides.\cite{Maczewsky2017, Mukherjee2017}

Beyond reproducing the topological behavior of time-independent systems, Floquet topological insulators also host phases which have no static analogue. Due to the periodic nature of the driving, it is possible to obtain phases with protected, gapless boundary states even though all bulk bands are topologically trivial. This is the case of so called anomalous topological phases,\cite{Kitagawa2010, Rudner2013, Quelle2017} in which bulk-boundary correspondence cannot be expressed solely in terms of bulk band topology, but requires other diagnostic tools. To date, there are proposals for anomalous versions of the quantum Hall effect,\cite{Kitagawa2010, Rudner2013} 2D TIs,\cite{Carpentier2015} topological superconductors,\cite{Jiang2011, Kundu2013, Reynoso2013, Thakurathi2013} WTIs,\cite{Lababidi2014, Ho2014} and also gapless systems, such as Weyl semimetals.\cite{Leykam2016, Zhou2016, Bomantara2016, Wang2016}

In this work, we extend the list of anomalous Floquet topological phases by considering mirror symmetry protected TCIs obtained by periodic driving. In these systems, protected surface states can appear even when the relevant topological index of the bulk bands, the mirror Chern number, takes only trivial values. In order to demonstrate the protection of boundary states, we introduce a new topological invariant. The latter is based on scattering theory, which provides a unified description capturing the topological properties of both static and driven systems.\cite{Fulga2016a} Extending this theory to the case of topological phases protected by reflection symmetry, similar to the recent work of Ref.~\onlinecite{Trifunovic2017}, allows to formulate an index that correctly determines the number of protected surface modes of both time-independent and Floquet systems, even in the anomalous phase.

Our construction of an anomalous Floquet TCI (AFTCI) is based on a coupled-layer approach,\cite{Trifunovic2016, Fulga2016b} and obtains mirror symmetric TCI phases from stacks of 2D quantum Hall-like systems. The number of protected boundary modes and their location in the surface Brillouin zone (BZ) can be controlled by adjusting the number of layers per unit cell and the inter-layer coupling, as in time-independent systems. Also similar to the static case, surface Dirac cones can be pinned to high symmetry points of the surface BZ and this pinning is captured by a $\mathbb{Z}_2$ scattering matrix invariant. We also highlight a feature which is unique to Floquet systems: the number of surface modes can be tuned by changing the driving protocol, such that doubling the driving period also doubles the number of protected surface states.

The rest of this work is organized as follows. In Section \ref{sec:layers} we briefly review the coupled-layer approach to TCIs, which obtains mirror symmetry protected topological phases from stacks of Chern insulators with alternating Chern numbers. In Section \ref{sec:aftci} we review the main building block of our construction, the 2D anomalous Floquet TI, and show that it straightforwardly leads to an AFTCI phase through the coupled-layer construction. In Section \ref{sec:cones} we demonstrate that surface states are topologically protected even for vanishing mirror Chern numbers by introducing scattering matrix invariants valid for reflection symmetric TCIs. We discuss how boundary modes can be manipulated by changing the number of layers in a unit cell, or by using larger driving periods in Section \ref{sec:manycones}. We conclude and discuss directions for future work in Section \ref{sec:conc}.

\section{Topological crystalline insulator as a layered system}
\label{sec:layers}

\begin{figure}[t]
 \includegraphics[width=0.7\columnwidth]{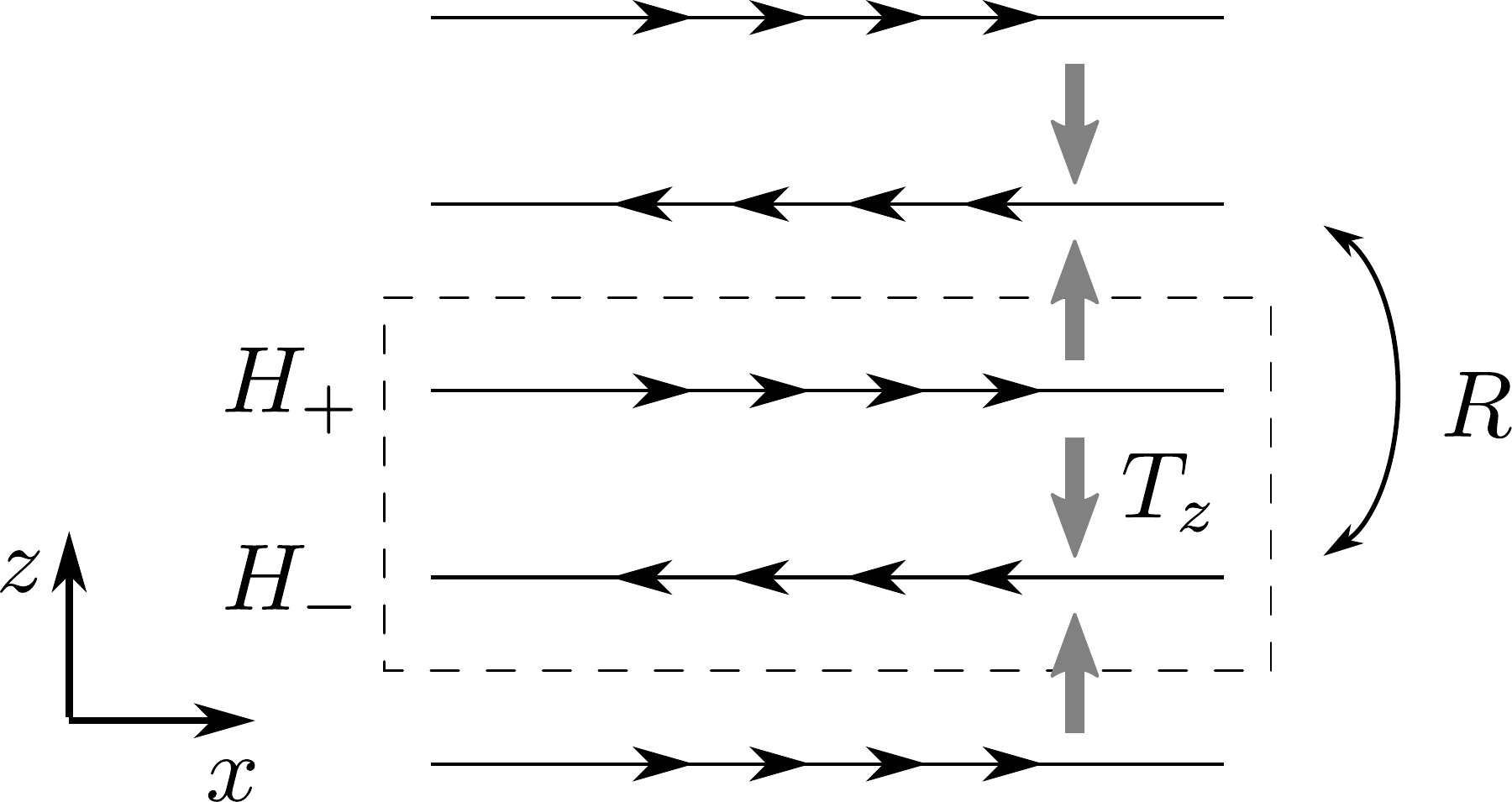}
 \caption{Side view of the TCI described by the Hamiltonian Eq.~\eqref{eq:htci}. The unit cell (dashed rectangle) is composed of two layers, each of which carries the same number of protected chiral edge modes (horizontal arrows). $H_+$ and $H_-$ have opposite Chern numbers, such that their edge states have opposite chirality. The layers are coupled as indicated by the vertical arrows: hopping along the arrows is described by the matrix $T_z$ and hopping in the opposite direction by $T^\dag_z$. Since each layer couples identically to its top and bottom neighbors, the model is symmetric with respect to reflection about one layer, which is described by the reflection operator $R$ of Eq.~\eqref{eq:reflop}.\label{fig:stack}}
\end{figure}

We begin by briefly reviewing the coupled-layer construction, a tunable toy model describing the topological properties of mirror symmetry protected TCIs.\cite{Fulga2016b} Consider a 3D system obtained as a stack of weakly coupled 2D gapped layers (see Fig.~\ref{fig:stack}). The layers alternate in the stacking direction $z$, being described by 2D Hamiltonians $H_+(k_x, k_y)$ and $H_-(k_x, k_y)$ with opposite values of their Chern numbers: $C_+=-C_-$. The latter constraint can be implemented, for instance, by ensuring that $+$ and $-$ layers are time-reversed partners, $H_\pm (k_x, k_y) = \Theta H_\mp (-k_x, -k_y) \Theta^{-1}$, with $\Theta$ the antiunitary time-reversal operator.

Because of their nonzero topological invariants, each layer hosts a total of $|C_+|=|C_-|$ protected chiral edge modes, which propagate in opposite directions for $H_+$ and $H_-$. In the presence of a nonzero inter-layer coupling, the counter-propagating edge modes of neighboring layers are allowed to backscatter and gap out. To produce a model of a 3D TCI, we choose a coupling term $T_z$ such that the stack is symmetric with respect to reflection about one layer. If a given layer couples identically to the neighbor above and the neighbor below, as shown in Fig.~\ref{fig:stack}, the resulting 3D Hamiltonian,

\begin{equation}\label{eq:htci}
 H_{\rm TCI} = \begin{pmatrix}
                H_+(k_x, k_y) & T_z^\pd (1 + e^{-i k_z}) \\
                T_z^\dag (1 + e^{i k_z}) & H_-(k_x, k_y) \\
               \end{pmatrix},
\end{equation}
obeys a reflection symmetry of the form
\begin{equation}\label{eq:reflsymh}
 H_{\rm TCI} (k_x, k_y, - k_z) = R(k_z) H_{\rm TCI} (k_x, k_y, k_z) R^\dag(k_z),
\end{equation}
with $R(k_z)$ a momentum-dependent unitary reflection operator
\begin{equation}\label{eq:reflop}
 R(k_z) = \begin{pmatrix}
           \mathbf{1} & \mathbf{0} \\
           \mathbf{0} & e^{-i k_z}\mathbf{1} \\
          \end{pmatrix}.
\end{equation}
The bold symbols appearing in each of the four entries in $R(k_z)$ signify that each block is a matrix of dimensions given by the number of orbitals in each layer, $H_\pm$. As such, $\mathbf{1}$ is the identity matrix and $\mathbf{0}$ is the zero matrix.

On the mirror symmetric $k_z=\pi$ plane of the BZ, the off-diagonal blocks of Eq.~\eqref{eq:htci} vanish. This means that the edge modes of the $H_+$ and $H_-$ sectors are uncoupled at $k_z=\pi$ due to mirror symmetry, and become coupled when $k_z\neq\pi$. For a system which is finite in the $y$-direction, the $k_x - k_z$ surface BZ will show a total of $|C_+|$ surface Dirac cones, which are protected by mirror symmetry and pinned to the $k_z=\pi$ line. The protection of surface modes can be expressed in terms of the topological invariant associated with mirror symmetry protected TCIs, the mirror Chern number
\begin{equation}\label{eq:cm}
 C_m = \frac{C_+ - C_-}{2}.
\end{equation}

In the following, we will show how to extend this model to the case of periodically driven systems.

\section{Anomalous Floquet TCI}
\label{sec:aftci}

In time-independent systems, the number of protected surface Dirac cones of the TCI Hamiltonian Eq.~\eqref{eq:htci} is dictated by the value of its mirror Chern number Eq.~\eqref{eq:cm} due to bulk-boundary correspondence. Similarly, each of the 2D layers forming the stack has a number of protected edge modes equal to its Chern number. If the layers were to be replaced with 2D systems hosting protected chiral edge states but with a Chern number equal to zero, then by Eq.~\eqref{eq:cm} the 3D stack would show protected surface Dirac cones for a vanishing mirror Chern number. As we will show, this scenario can be realized with the help of periodic driving, when each of the layers forms an anomalous Floquet topological phase.\cite{Kitagawa2010}

\subsection{Two-dimensional Floquet layer}

\begin{figure}[tb]
 \includegraphics[width=\columnwidth]{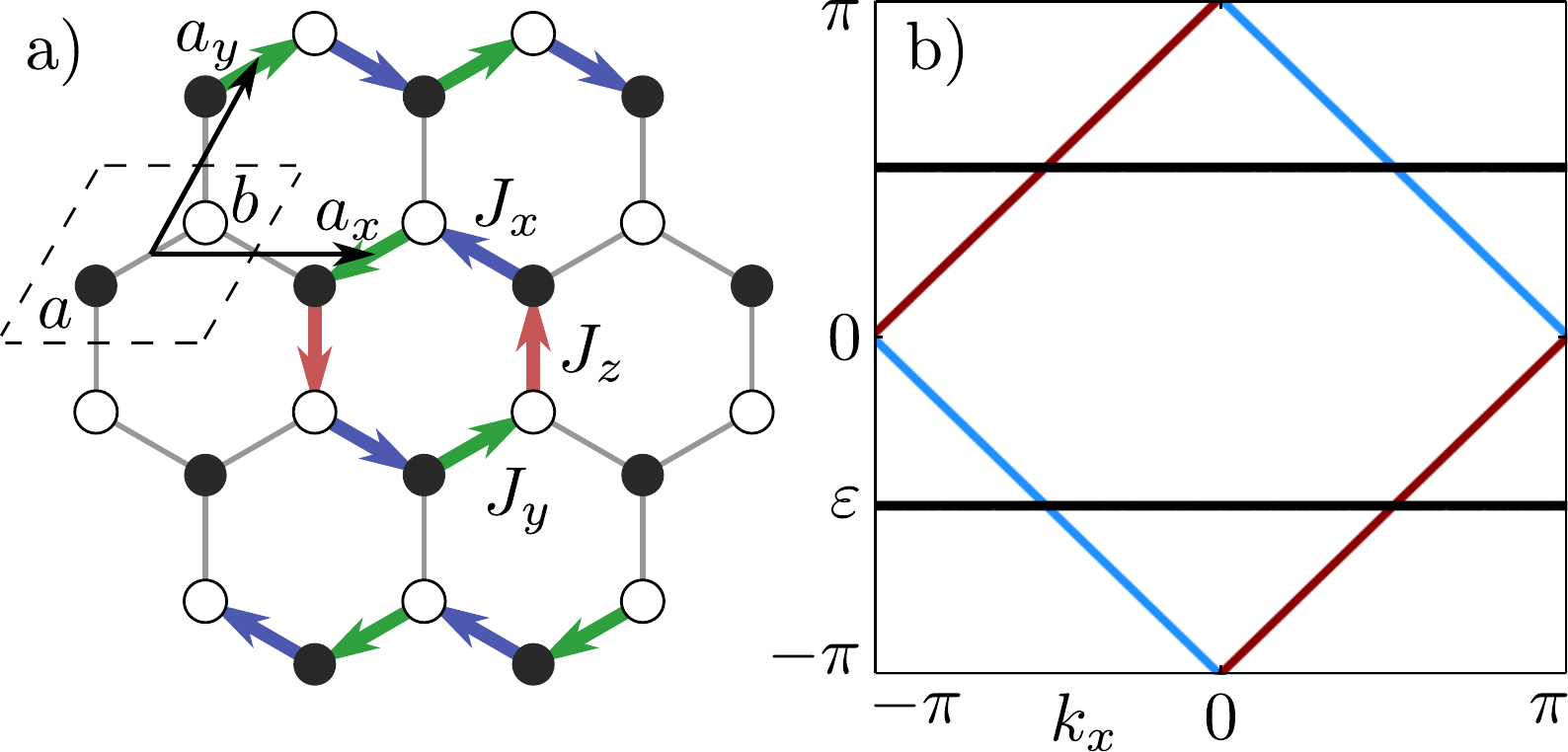}
 \caption{Panel a: Setup and driving protocol of the Hamiltonian Eq.~\eqref{eq:hhex}. The model describes spinless fermions on a hexagonal lattice (unit cell marked by dashed lines, sublattices $a$ and $b$, Bravais vectors $a_x$ and $a_y$). At any given time, there is only one nonzero hopping $J_x$, $J_y$, or $J_z$, shown as blue, green, and red arrows, respectively. When $JT/3=\pi/2$, particles in the bulk return to their original position after a time $2T$, while those on the edge propagate unidirectionally. Panel b: Bandstructure obtained by diagonalizing the Floquet operator Eq.~\eqref{eq:fl3steps} in an infinite strip geometry (infinite in $a_x$, 10 unit cells along $a_y$), setting $JT/3=\pi/2$. The spectrum consists of two flat bulk bands (black), as well as one chiral mode at each boundary of the system (red, blue). The edge modes wind in the BZ along both momentum and quasienergy.\label{fig:hex}}
\end{figure}

Our main building block is a layer described by a 2D tight-binding model of spinless fermions on a hexagonal lattice (see Fig.~\ref{fig:hex}a). The Hamiltonian reads
\begin{equation}\label{eq:hhex}
 H = \sum_{\bf k} \begin{pmatrix}
                   c^\dag_{a, {\bf k}} & c^\dag_{b, {\bf k}} \\
                  \end{pmatrix}
{\cal H}({\bf k}) \begin{pmatrix}
                   c^\pd_{a, {\bf k}} \\ c^\pd_{b, {\bf k}} \\
                  \end{pmatrix},
\end{equation}
with
\begin{equation}\label{eq:hlayer}
\begin{split}
\cal{H}({\bf{k}}) =& [J_x \cos(k_x) + J_z \cos(k_y) + J_y]\sigma_x - \\
& [J_x \sin(k_x) + J_z \sin(k_y)]\sigma_y
\end{split}
\end{equation}

In Eqs.~\eqref{eq:hhex} and \eqref{eq:hlayer}, ${\bf k} = (k_x, k_y)$ are the two momenta, $c^{(\dag)}$ are fermionic creation and annihilation operators, the Pauli matrices $\sigma_i$ parametrize the degree of freedom associated to the two sublattices labeled $a$ and $b$, and $J_x$, $J_y$, and $J_z$ denote the nearest neighbor hopping amplitudes in three different directions (see Fig.~\ref{fig:hex}a).

The Hamiltonian Eq.~\eqref{eq:hlayer} can describe an anomalous topological phase if the hopping amplitudes $J_i$ ($i=x,y,z$) are varied periodically in time: $J_i(t) = J_i(t+T)$, with $T$ the driving period. We chose a stroboscopic driving scheme in which the hoppings are successively turned on, such that only one of the $J_i$ is nonzero at a given time. The driving protocol proceeds as: 
\begin{enumerate}
 \item $J_x=J$, $J_{y,z}=0$ for $nT<t\leq nT+T/3$,
 \item $J_y=J$, $J_{z,x}=0$ for $nT+T/3<t\leq nT+2T/3$,
 \item $J_z=J$, $J_{x,y}=0$ for $nT+2T/3<t\leq (n+1)T$,
\end{enumerate}
with $n\in{\mathbb Z}$. Since energy is no longer conserved, we describe the behavior of the system in terms of the unitary time-evolution operator over one period, called the Floquet operator
\begin{equation}\label{eq:floquet}
 {\cal F}({\bf k}) = {\cal T} \exp \left( -i \int_0^T dt\, {\cal H}({\bf k}, t) \right),
\end{equation}
where ${\cal T}$ denotes time ordering and we have set $\hbar=1$. For the driving protocol written above, the Floquet operator reads
\begin{equation}\label{eq:fl3steps}
 {\cal F}({\bf k}) = e^{-i{\cal H}_3({\bf k})T/3}e^{-i{\cal H}_2({\bf k})T/3}e^{-i{\cal H}_1({\bf k})T/3}
\end{equation}
with ${\cal H}_{1,2,3}$ the Hamiltonian of Eq.~\eqref{eq:hlayer} during each of the three steps of the driving protocol.

The eigenvalues and eigenvectors of the Floquet operator Eq.~\eqref{eq:fl3steps} are analogous to the energies and wave functions of time-independent systems. When translation symmetry is preserved, momentum is a good quantum number, such that diagonalizing 
${\cal F}({\bf k})|\psi_{\bf k}\rangle = \exp (-i \varepsilon_{\bf k} T)|\psi_{\bf k}\rangle$ yields the Floquet states $|\psi_{\bf k}\rangle$ and the associated quasienergies $\varepsilon_{\bf k}$. Because of the unitarity of ${\cal F}$ however, the $\varepsilon_{\bf k}$ are only defined modulo $2\pi/T$, such that the Brillouin zone is not only periodic in momentum, but also in quasienergy. This feature is unique to periodically-driven systems and enables the formation of an anomalous topological phase, as we discuss below.

At each point during the time evolution the system is described by a set of decoupled dimers, since only one of the hopping terms $J_i$ is nonzero, taking a value $J$. Choosing this value such that $JT/3=\pi/2$ ensures that a particle on a given site is transferred with unit probability to the neighboring site during one step of the driving protocol. As shown in Fig.~\ref{fig:hex}a, this has two consequences. First, a particle initially located on a bulk site will return to its initial position after two driving periods, leading to the formation of two dispersionless Floquet bulk bands. Second, a particle on an edge site will keep propagating unidirectionally along the edge, leading to the formation of chiral Floquet edge modes.
In Fig.~\ref{fig:hex}b we plot the bandstructure of the system in an infinite strip geometry using the {\sc kwant} code \cite{Groth2014, SuppMat}. There are flat bulk bands at quasienergies $\varepsilon=\pm\pi/2$, as well as dispersive edge modes which wind around the BZ in both momentum and quasienergy. 

The position of the edge modes in the bandstructure of Fig.~\ref{fig:hex}b is constrained by particle-hole symmetry. Due to the form of the Hamiltonian Eq.~\eqref{eq:hlayer}, the Floquet operator Eq.~\eqref{eq:fl3steps} obeys a particle-hole symmetry of the form ${\cal F}({\bf k})={\cal F}^*(-{\bf k})$, such that for every Floquet eigenstate at quasienergy $\varepsilon$ and momentum ${\bf k}$ there exists another state at $-\varepsilon$ and $-{\bf k}$, similar to the behavior of static systems. Since there is only one chiral mode on each boundary of the system, the edge state must exist at points which are symmetric under $\varepsilon\to-\varepsilon$ and $k_x\to-k_x$. This pins the edge modes of Fig.~\ref{fig:hex}b to $\varepsilon=0$, $k_x=\pi$, as well as to $\varepsilon=\pi$ and $k_x=0$.

At quasienergies for which no bulk states are available, the only backscattering process involves tunneling from one edge to the other, since the edge modes are chiral. Such a process has an exponentially small amplitude in the inter-edge separation, so the edge modes have the same level of protection as in the quantum Hall effect. As shown in Refs.~\onlinecite{Kitagawa2010, Rudner2013} however, the Chern numbers associated with the two bulk bands vanish. This can be directly seen in Fig.~\ref{fig:hex}b by noticing that for a given band, the Chern number counts the net difference between the number of edge modes in the gap above the band and the number of edge modes in the gap below it. Since edge modes wind around the BZ in quasienergy, this difference is always zero.

\subsection{Stacked Floquet TCI}

Having obtained a model for a 2D system with protected edge modes but trivial bulk bands, we can straightforwardly apply the coupled layer framework to obtain an anomalous Floquet TCI. We replace both diagonal blocks of the TCI Hamiltonian Eq.~\eqref{eq:htci} with the hexagonal model of Eq.~\eqref{eq:hlayer}. Now both $H_+$ and $H_-$ will form 2D anomalous Floquet topological phases when periodic driving is included. To ensure that adjacent layers have edge modes moving in opposite directions, we use the three-step driving protocol shown above for the $H_+$ layers ($1\to2\to3$), but the inverted sequence of steps for the $H_-$ layers ($3\to2\to1$). For a single layer, such an inverted sequence amounts to applying a time-reversal transformation, which on the level of the Floquet operator Eq.~\eqref{eq:fl3steps} means taking the transpose: ${\cal F}_-^T = {\cal F}^\pd_+ \equiv {\cal F}$, with the superscript $T$ denoting transposition. The change of edge state chirality when inverting the driving protocol can also be seen by considering the motion of particles on a finite lattice, as done in Fig.~\ref{fig:hex}a. Finally, we introduce a time-independent inter-layer coupling that connects sites in adjacent layers and in the same sublattice. The three-step AFTCI Floquet operator is similar to Eq.~\eqref{eq:fl3steps},
\begin{equation}\label{eq:faftci}
 {\cal F}_{\rm AFTCI}  = \prod_{j=3,2,1}\exp\left[-i\frac{T}{3}\begin{pmatrix}
                             {\cal H}_j & A \\
                             A^\dag & {\cal H}_{4-j} \\
                            \end{pmatrix}\right]
\end{equation}
with $A=t_z\sigma_0(1 + e^{-ik_z})$ and ${\cal H}_{j}$ as in Eq.~\eqref{eq:fl3steps}.

\begin{figure}[tb]
 \includegraphics[width=\columnwidth]{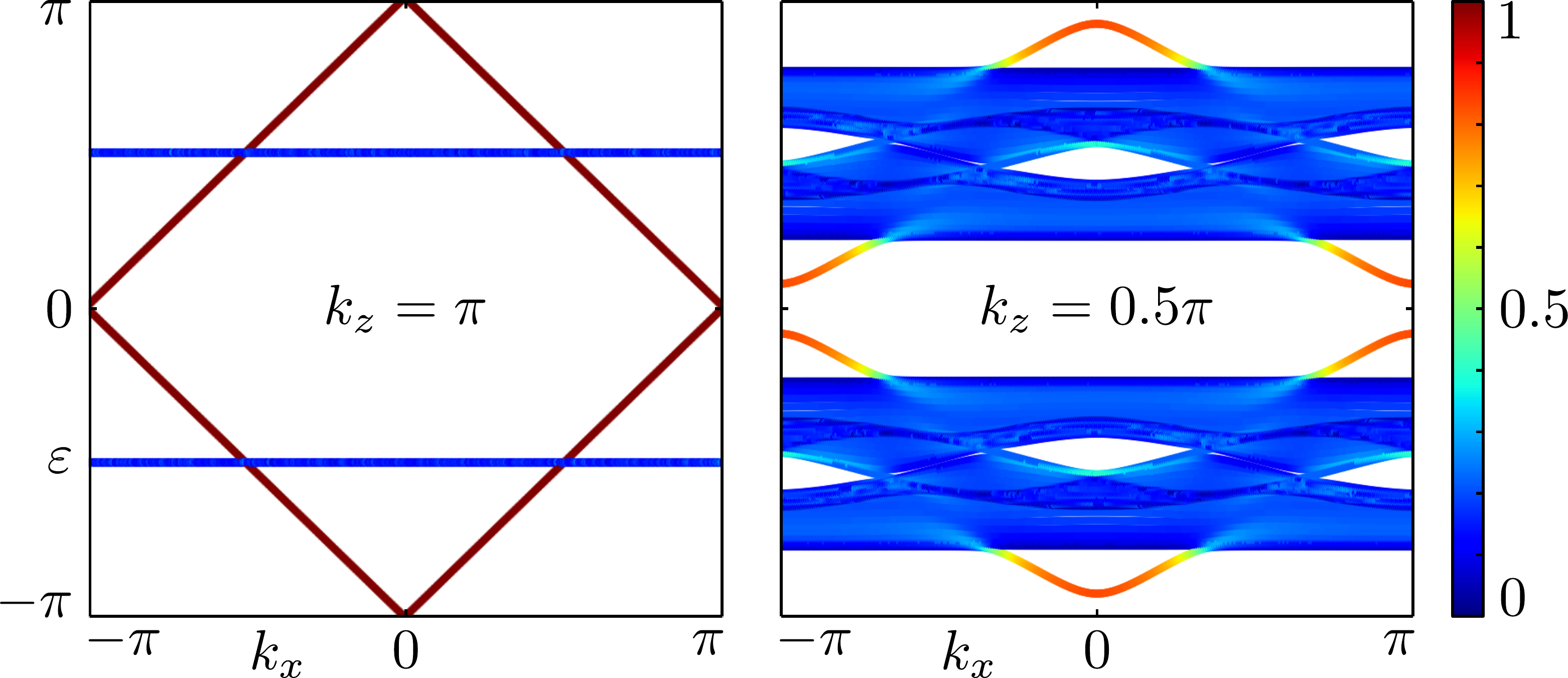}
 \caption{Bandstructure of the anomalous Floquet TCI in an infinite slab geometry, with hard wall boundaries in the $a_y$ direction and a thickness of 20 unit cells. The color scale represents the eigenstate intensity on the first and last two unit cells from the boundary. The layers' edge modes decouple at the mirror symmetric momentum $k_z=\pi$ (left), while they backscatter and gap out away from this point (right, $k_z=\pi/2$). There are two protected Dirac cones on each surface (see Appendix \ref{app:DCpos}), one at $\varepsilon=0$ and one at $\varepsilon=\pm\pi$.\label{fig:tci1}}
\end{figure}

We set $t_zT=0.6$ and $JT/3=\pi/2$ throughout the following and plot in Fig.~\ref{fig:tci1} the bandstructure of the system 
in an infinite slab geometry, with hard wall boundary conditions only in the $a_y$ direction. As before, on the mirror symmetric $k_z=\pi$ plane of the BZ, the layers are decoupled, so the bandstructure is given by Fig.~\ref{fig:hex}b superimposed with its time-reversed partner. Away from the mirror symmetric plane, the edge modes gap out, such that there are two Dirac cones on each surface of the slab, located at $k_z=\pi$, one in the gap at $\varepsilon=0$ and one in the gap at $\varepsilon=\pi$. By Eq.~\eqref{eq:cm}, the mirror Chern numbers associated to the bands vanish, since each individual layer has trivial Chern numbers, $C_+=C_-=0$. Similar to the 2D case, the anomalous nature of the Floquet TCI phase implies that each bulk band has an equal number of surface Dirac cones in the gap above it and in the gap below it.

The position of the surface Dirac cones is pinned to $k_z=\pi$ due to the fact that the Floquet operator of the 3D system obeys the same mirror symmetry as in the static case,
\begin{equation}\label{eq:AFTCI_mirror}
\begin{split}
 {\cal F}_{\rm AFTCI}(k_x, k_y, -k_z) &= \\
 R(k_z)&{\cal F}_{\rm AFTCI}(k_x, k_y, k_z)R^\dag(k_z),
\end{split}
\end{equation}
with $R(k_z)$ given by Eq.~\eqref{eq:reflop}. Further, the Dirac cones are also constrained to appear at high symmetry points in the $k_x$ direction. Similar to the 2D system, this is due to a particle-hole symmetry of the layered system,
\begin{equation}\label{eq:AFTCI_phs}
 {\cal F}_{\rm AFTCI}({\bf k}) = 
 \begin{pmatrix}
  \mathbf{1} & \mathbf{0} \\
  \mathbf{0} & -\mathbf{1} \\
 \end{pmatrix}
          {\cal F}^*_{\rm AFTCI}(-{\bf k})
 \begin{pmatrix}
  \mathbf{1} & \mathbf{0} \\
  \mathbf{0} & -\mathbf{1} \\
 \end{pmatrix},
\end{equation}
where ${\bf k}=(k_x,k_y,k_z)$, and $\mathbf{0},\mathbf{1}$ are $2\times2$ matrices acting on each of the layers of Eq.~\eqref{eq:faftci}. Due to Eq.~\eqref{eq:AFTCI_phs}, for every state at quasienergy $\varepsilon$ and momentum ${\bf k}$ there must also exist a state at $-\varepsilon$ and $-{\bf k}$. As such, when the number of surface Dirac cones in a gap is odd, they must appear at high-symmetry points. This is shown in Fig.~\ref{fig:tci1}, where one of the Dirac cones appears at $\varepsilon=0$ and $k_x=\pi$, whereas the other sits at $\varepsilon=\pi$ and $k_x=0$. However, if the total number of surface Dirac cones in a given gap is even, then particle-hole symmetry does not constrain their precise position.

\section{Topological invariants}
\label{sec:cones}

Both in the 2D anomalous Floquet TI as well as in the 3D AFTCI, the topology of the bulk bands does not correctly capture the number of protected boundary modes. To circumvent this issue and demonstrate topological protection, Ref.~\onlinecite{Rudner2013} have defined new topological indices for the 2D system, which take into account not just the Floquet operator, but the full time-evolution operator throughout the driving period. In this Section we will use a different approach, expressing the topological indices of the AFTCI with the help of scattering theory.

\subsection{Two-dimensional invariants}

Consider a 2D system described by the Floquet operator Eq.~\eqref{eq:fl3steps}, in a ribbon geometry, which is infinite along $a_x$ and contains $W$ unit cells in the $a_y$ direction, labeled by an integer $1\leq n_y \leq W$. We define a scattering matrix by studying the time evolution of the system in the presence of absorbing terminals introduced on the last unit cell of each boundary, at $n_y= 1$ and $n_y=W$.\cite{Fulga2016a}
After each full period of the time evolution, the terminals project out any wave function component located on the absorbing sites, leading to an expression for the scattering matrix
\begin{equation}\label{eq:sfl}
 S(k_x,\varepsilon) = P \left[ 1-e^{i\varepsilon}{\cal F}(k_x)(1-P^TP) \right]^{-1}e^{i\varepsilon}{\cal F}(k_x)P^T,
\end{equation}
where the superscript $T$ denotes transposition, ${\cal F}(k_x)$ is the $2W\times2W$ Floquet operator of the ribbon, and $P$ is the $4\times2W$ projector onto the two terminals,
\begin{equation}\label{eq:projop}
P = \left\{
\begin{array}{ll}
      1 & n_y =1 \text{ or } n_y =W\\
      0 & \text{otherwise} \\
\end{array} 
\right.  
\end{equation}
 
As shown in Ref.~\onlinecite{Fulga2016a}, the two terminal scattering matrix Eq.~\eqref{eq:sfl} takes the form
\begin{equation}\label{eq:smatblocks}
	S(k_x, \varepsilon)=\begin{pmatrix}
    r(k_x, \varepsilon) & t(k_x, \varepsilon) \\
    t'(k_x, \varepsilon) & r'(k_x, \varepsilon)
\end{pmatrix},
\end{equation}
where $t^{(\prime)}$ and $r^{(\prime)}$ are blocks containing the probability amplitudes for a particle to be transmitted from one terminal to the other, or to be reflected back into the same terminal, respectively. Each block has a size $2\times2$, encoding the degree of freedom associated to the $a$ and $b$ sublattices. 

\begin{figure}[tb]
 \includegraphics[width=\columnwidth]{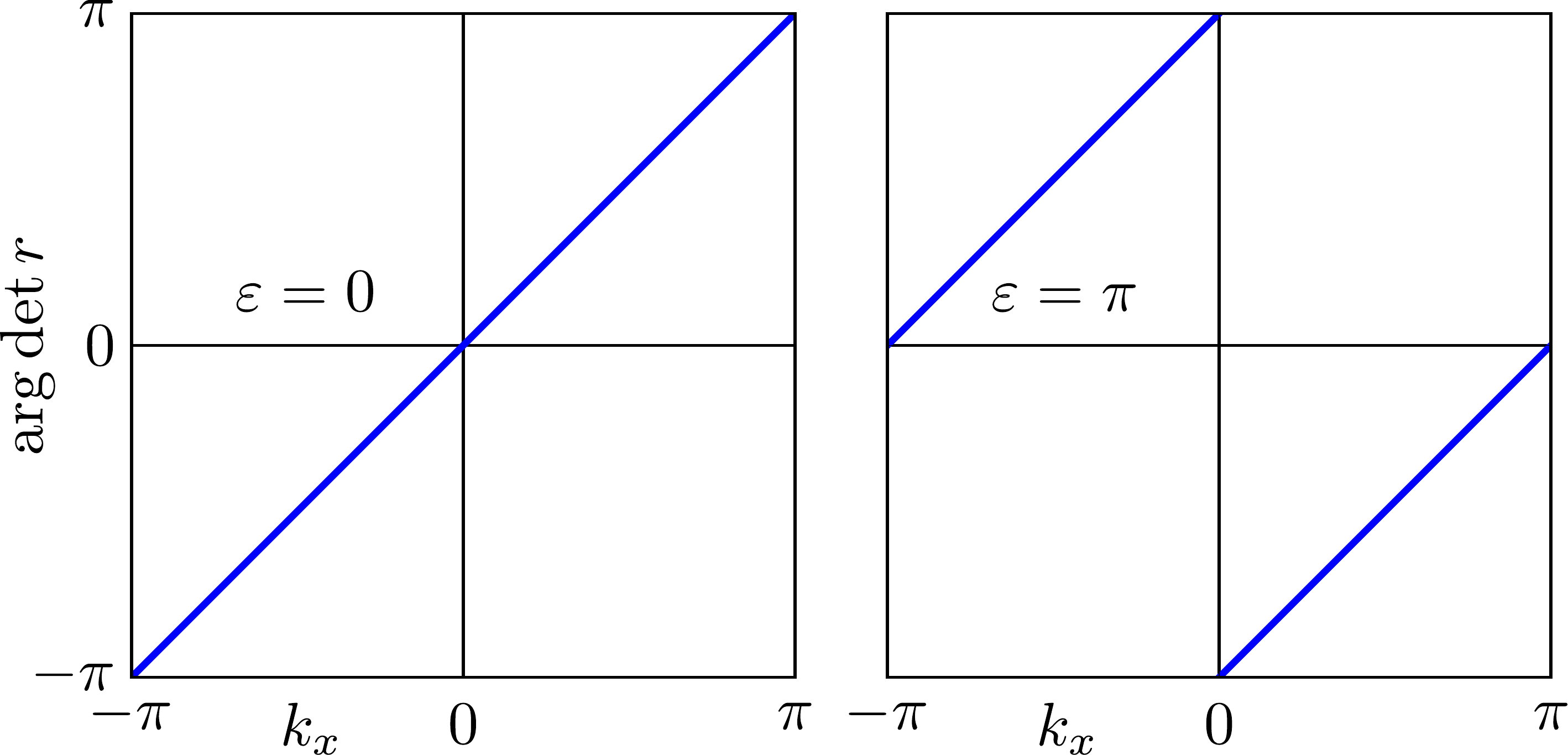}
 \caption{The phase of the reflection block determinant, ${\rm arg}\det r$ is computed for an infinite ribbon of the 2D model Eq.~\eqref{eq:fl3steps} ($W=6$ unit cells in the $a_y$ direction) and plotted as a function of momentum $k_x$. In the left and right panels $\varepsilon=0$ and $\varepsilon=\pi$, respectively. At both values of quasienergy, the phase winds in the positive direction as a function of momentum, signaling a non-trivial strong topological invariant ${\cal W}(\varepsilon=0)={\cal W}(\varepsilon=\pi)=1$. Further, at particle-hole symmetric points $k_x=0,\pi$, the phase determines the weak topological invariants, $\nu(k_x, \varepsilon)$. These take the values $\nu(0,0)=\nu(\pi,\pi)=0$ and $\nu(0,\pi)=\nu(\pi,0)=1$, such that both the strong and the weak indices are consistent with the bandstructure shown in Fig.~\ref{fig:hex}. \label{fig:inv2d}}
\end{figure}

At any quasienergy for which the bulk is insulating, the scattering matrix Eq.~\eqref{eq:smatblocks} enables to define a topological invariant counting the net number of chiral modes at the boundary of the system. The index is the winding number of the reflection block determinant,\cite{Braeunlich2009, Fulga2012, Fulga2016a}
\begin{equation}\label{eq:s_inv}
{\cal W}({\varepsilon}) = \frac{1}{2\pi i}\int_0^{2\pi} dk_x\,\frac{d}{dk_x}\log\,\det\,r(k_x, \varepsilon).
\end{equation}
For the 2D system shown in Fig.~\ref{fig:hex}, we find ${\cal W}(\varepsilon=0)={\cal W}(\varepsilon=\pi)=1$ (see Fig.~\ref{fig:inv2d}), consistent with the presence of a single chiral mode on each boundary. Further, the pinning of the edge modes to high-symmetry points in the surface BZ is captured by a different topological invariant. The particle-hole symmetry of the 2D Floquet operator, ${\cal F}({\bf k})={\cal F}^*(-{\bf k})$, translates into a particle-hole symmetry of the scattering matrix, 
\begin{equation}\label{eq:smat_phs_2d}
 S(k_x, \varepsilon) = S^*(-k_x, -\varepsilon)
\end{equation}
due to Eqs.~\eqref{eq:sfl} and \eqref{eq:smatblocks}. This enables to define weak topological invariants associated to the high-symmetry points $k_x=0,\pi$ and particle-hole symmetric quasienergies $\varepsilon=0,\pi$:
\begin{equation}\label{eq:s_weak_inv}
 (-1)^{\nu(k_x, \varepsilon)} = {\rm sign} \det r(k_x, \varepsilon),
\end{equation}
such that the $\mathbb{Z}_2$ index $\nu(k_x, \varepsilon)$ takes the values 0 or 1. As shown in Ref.~\onlinecite{Fulga2016a}, the invariants Eq.~\eqref{eq:s_weak_inv} count the parity of the number of edge modes present at particle-hole symmetric points of the edge BZ, similar to the behavior of time-independent weak topological superconductors.\cite{Diez2015} Since the phase of $\det r$ plotted in Fig.~\ref{fig:inv2d} is equal to $\pi$ both for $\varepsilon=0$ and $k_x=\pi$ as well as for $\varepsilon=\pi$ and $k_x=0$, we find $\nu(0,\pi)=\nu(\pi, 0)=1$, indicating that there are an odd number of edge modes present at those high-symmetry points (see Fig.~\ref{fig:hex}). On the other hand, $\nu(0,0)=\nu(\pi, \pi)=0$, so there are an even number of edge modes (in this case zero) at $k_x=\varepsilon=0,\pi$.

Nonzero values of the strong and weak topological invariants Eq.~\eqref{eq:s_inv} and \eqref{eq:s_weak_inv} imply the presence of a robust topological phase hosting protected edge modes, which can be seen as follows. First, note that the winding number of the phase of $\det r$ (which is the same as the number of chiral edge modes), cannot change unless $\det r=0$, at which point the bulk of the system becomes conducting. As such, a phase with ${\cal W}(\varepsilon)\neq 0$ will host the same number of edge modes no matter how the system is perturbed, as long as the perturbation does not close the Floquet bulk gap at the quasienergy $\varepsilon$. This includes changing the strength of the hoppings, adding longer range coupling terms to the system, or adding on-site perturbations. Similarly, as long as the bulk remains insulating and the particle-hole symmetry constraint of Eq.~\eqref{eq:smat_phs_2d} is obeyed, the weak invariant Eq.~\eqref{eq:s_weak_inv} cannot change, which means that chiral modes remain pinned to the high-symmetry points $k_x=0,\pi$ even if the system is perturbed.

\subsection{Mirror invariants of the TCI}

\begin{figure}[tb]
 \includegraphics[width=\columnwidth]{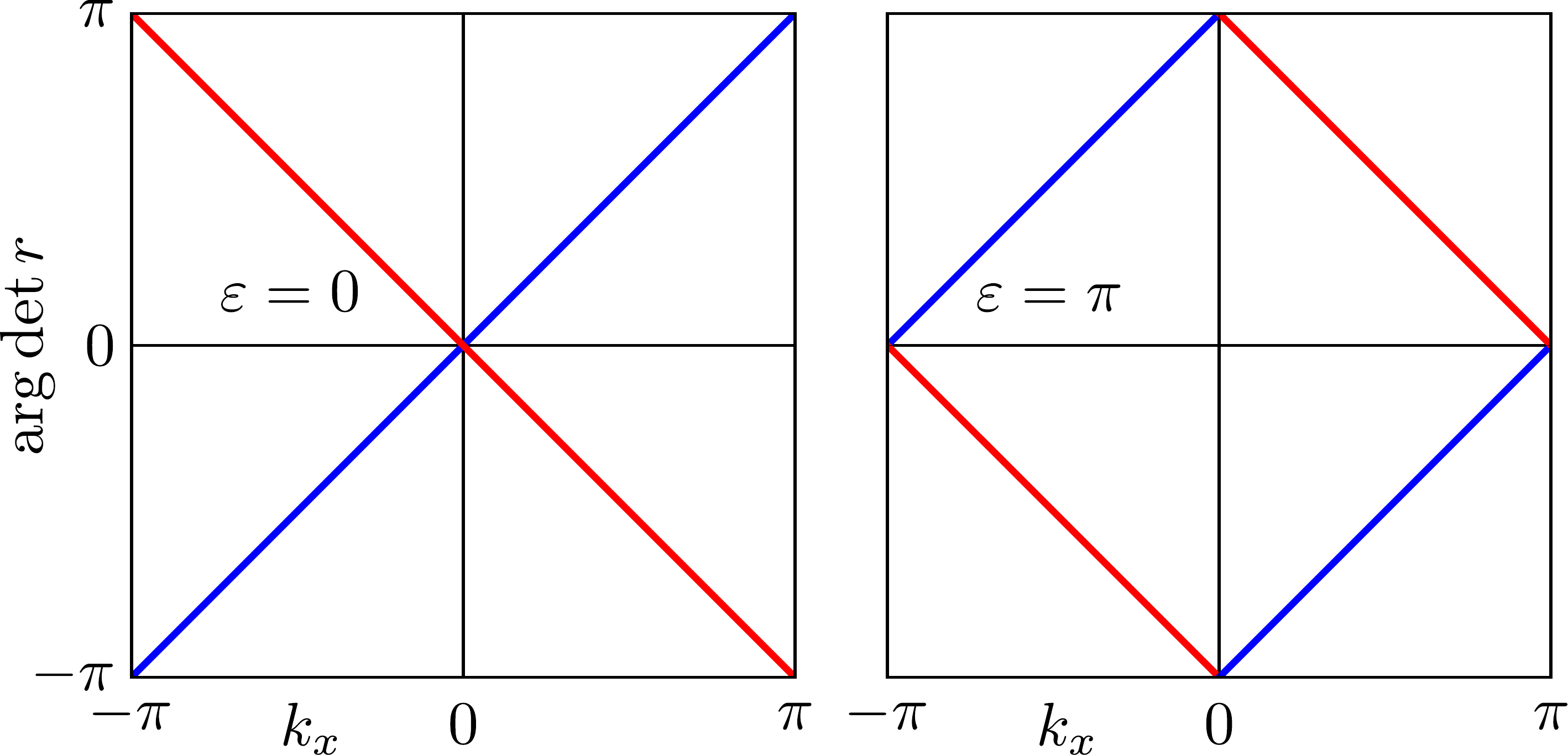}
 \caption{The phase of each block of the reflection matrix, $r_\pm$ plotted as a function of momentum $k_x$. We use an infinite slab of the AFTCI model Eq.~\eqref{eq:faftci} in which $k_x$ and $k_z$ are good quantum numbers and which consists of $W=6$ unit cells in the $a_y$ direction. The blue and red colors indicate blocks $r_+$ and $r_-$, respectively, whereas left and right panels correspond to $\varepsilon=0$ and $\varepsilon=\pi$. The phases of the two reflection blocks wind once, but in opposite directions, such that ${\cal W}_m=1$, consistent with the number of surface Dirac cones shown in Fig.~\ref{fig:tci1}. Similarly, the weak mirror invariants $\nu_m$ of Eq.~\eqref{eq:mirror_nu} are non-trivial only at $\varepsilon=0$, $k_x=\pi$ and $\varepsilon=\pi$, $k_x=0$, showing that there exist an odd number of surface Dirac cones at those points of the BZ.\label{fig:inv_tci_1}}
\end{figure}

Applying the layered construction to obtain time-independent TCIs allows to trace the topological invariant of the 3D system, the mirror Chern number, to the 2D invariants characterizing the layers, the Chern numbers $C_+$ and $C_-$. In a similar fashion, for the 3D Floquet system described by Eq.~\eqref{eq:faftci}, it is possible to define a strong and weak mirror indexes starting from the 2D scattering matrix invariants Eq.~\eqref{eq:s_inv} and \eqref{eq:s_weak_inv}. To this end, we consider the layered model in the infinite slab geometry used in Fig.~\ref{fig:tci1}, such that $k_x$ and $k_z$ remain good quantum numbers. Repeating the procedure of Eqs.~\eqref{eq:sfl} and \eqref{eq:projop}, we define absorbing terminals on the boundaries of the slab and obtain a Floquet scattering matrix $S(k_x,k_z,\varepsilon)$ with the same block structure as Eq.~\eqref{eq:smatblocks}. The mirror and particle-hole symmetries of ${\cal F}_{\rm AFTCI}$ then translate into constraints for the reflection block of the scattering matrix, $r(k_x,k_z,\varepsilon)$. Due to Eq.~\eqref{eq:AFTCI_mirror} we find
\begin{equation}\label{eq:rblock_mirror}
 r(k_x,k_z,\varepsilon)=R(k_z)r(k_x,-k_z,\varepsilon)R^\dag(k_z),
\end{equation}
whereas the combination of mirror and particle-hole symmetries, Eqs.~\eqref{eq:AFTCI_mirror} and \eqref{eq:AFTCI_phs} leads to
\begin{equation}\label{eq:rblock_phs}
 r(k_x,k_z=\pi,\varepsilon)=
 r^*(-k_x,k_z=\pi,-\varepsilon)
\end{equation}
on the mirror plane. Due to mirror symmetry, the reflection block of the scattering matrix becomes block-diagonal on the mirror invariant $k_z=\pi$ plane of the BZ:
\begin{equation}\label{eq:rblock_diag}
 r(k_x,\pi,\varepsilon)=
 \begin{pmatrix}
  r_+(k_x,\varepsilon) & \mathbf{0} \\
  \mathbf{0} & r_-(k_x,\varepsilon) \\
 \end{pmatrix}.
\end{equation}
This can also be understood by noticing that the Floquet operator describes decoupled layers on the mirror plane, such that a particle initially located on an $H_+$ layer cannot be reflected out of an $H_-$ layer. The Eq.~\eqref{eq:rblock_diag} enables us to generalize the topological invariant ${\cal W}(\varepsilon)$ to a \textit{mirror winding number} characterizing the 3D AFTCI. We define
\begin{equation}\label{eq:mirror_W}
 {\cal W}_m(\varepsilon) = \frac{{\cal W}_+(\varepsilon)-{\cal W}_-(\varepsilon)}{2},
\end{equation}
similar to the mirror Chern number of Eq.~\eqref{eq:cm}, where the winding numbers ${\cal W}_\pm$ describe the topological invariants of each layer, computed by applying Eq.~\eqref{eq:s_inv} to the $r_\pm$ blocks of the reflection matrix. Notice that since adjacent layers are time-reversed partners, ${\cal W}_+(\varepsilon)=-{\cal W}_-(\varepsilon)$, which implies that ${\cal W}_m(\varepsilon)$ is an integer. In Fig.~\ref{fig:inv_tci_1} we show the winding of the phase of $\det r_\pm$, leading to a value of ${\cal W}_{m}=1$ both for the gap at $\varepsilon=0$ and for that at $\varepsilon=\pi$. As such, the mirror topological invariant correctly captures the number of surface Dirac cones of ${\cal F}_{\rm AFTCI}$ (see Fig.~\ref{fig:tci1}).

In a similar fashion, we can introduce a \textit{mirror weak index}
\begin{equation}\label{eq:mirror_nu}
 \nu_m(k_x, \varepsilon) = \frac{\nu_+(k_x, \varepsilon)+\nu_-(k_x, \varepsilon)}{2},
\end{equation}
since the reflection matrix is real on the mirror plane [Eq.~\eqref{eq:rblock_phs}], allowing to define $(-1)^{\nu_\pm(k_x,\varepsilon)} = {\rm sign}\det r_\pm (k_x,\varepsilon)$.Again, due to the fact that neighboring layers are related by time-reversal symmetry, $\nu_+(k_x, \varepsilon)=\nu_-(k_x, \varepsilon)$, such that $\nu_m(k_x, \varepsilon)$ is a $\mathbb{Z}_2$ index, taking the values 0 or 1. A nonzero value of this quantity implies the presence of an odd number of surface Dirac cones at the high-symmetry point $k_x=0$ or $\pi$ and corresponding particle-hole symmetric quasienergy $\varepsilon=0$ or $\pi$. As can be seen from Fig.~\ref{fig:inv_tci_1}, the only non-trivial invariants are $\nu_m(\pi,0)=\nu_m(0,\pi)=1$, correctly capturing the position of the surface Dirac cones.

Similar to the two-dimensional case discussed previously, nonzero mirror invariants Eq.~\eqref{eq:mirror_W} and \eqref{eq:mirror_nu} describe a robust Floquet TCI hosting topologically protected surface Dirac cones. As long as mirror symmetry constraint of Eq.~\eqref{eq:rblock_mirror} is preserved and the bulk remains insulating on the mirror invariant plane ($\det r_{\pm}\neq0$), neither the mirror winding number ${\cal W}_m(\varepsilon)$ nor the number of surface Dirac cones in the gap at $\varepsilon$ can change. Further, if in addition particle-hole symmetry [Eq.~\eqref{eq:rblock_phs}] remains valid, then the weak index $\nu_m$ as well as the parity of the number of surface Dirac cones at high-symmetry points remains constant. The resulting Floquet TCI is therefore robust to any symmetry preserving perturbations which do not close the bulk gap, such as changing the in-plane hopping $J$ or the inter-layer coupling $t_z$, even if the latter becomes time-dependent.

\section{Tunable surface Dirac cones}
\label{sec:manycones}

Having established both a model for an anomalous Floquet TCI and the topological invariants expressing its bulk-surface correspondence, we explore in the following the layered construction in more detail. In particular, we show that the number of Dirac cones may be tuned by increasing the size of the unit cell, but also by keeping the number of layers fixed and varying instead the driving protocol. In addition, we show that for larger numbers of surface states, the strong and weak mirror invariants do not always capture the exact position of surface Dirac cones, but only their number and their parity at high symmetry points.

\begin{figure}[tb]
 \includegraphics[width=\columnwidth]{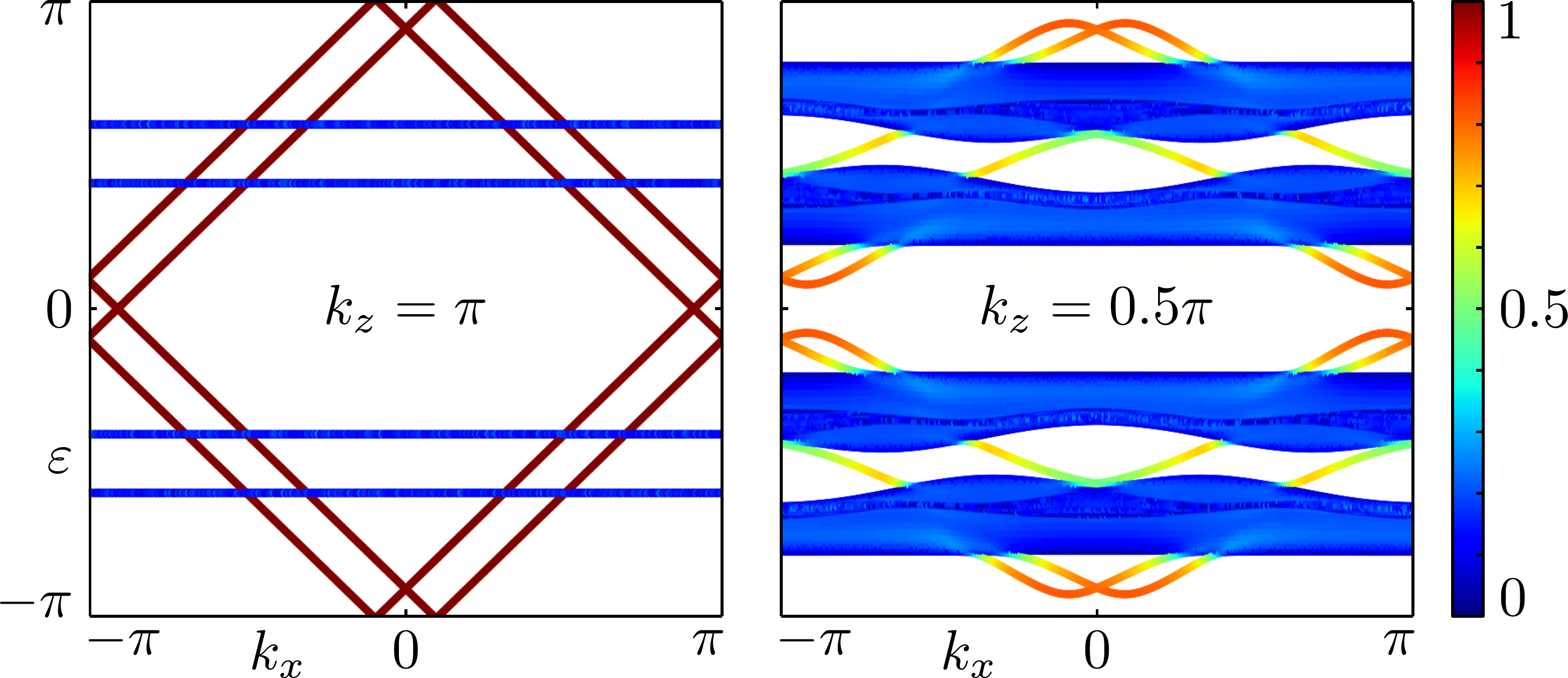}
 \caption{Same as Fig.~\ref{fig:tci1}, but for the Floquet operator ${\cal F}_2$ [Eq.~\eqref{eq:fl_doubled}], which contains four layers per unit cell. There are now four surface Dirac cones on each surface, two at $\varepsilon=0$ and two at $\varepsilon=\pi$. They are located away from the high symmetry points $k_x=0,\pi$ due to vanishing weak mirror invariants, but their position is still symmetric with respect to $k_x\to-k_x$ due to particle-hole symmetry.\label{fig:tcipm2}}
\end{figure}

To obtain a system with a mirror winding number ${\cal W}_m=2$, we can simply double the size of the Floquet operator ${\cal F}_{\rm AFTCI}$. We consider a system containing four layers per unit cell
\begin{equation}\label{eq:fl_doubled}
 {\cal F}_2 = \prod_{j=3,2,1} \exp \left[ -i\frac{T}{3}
 \begin{pmatrix}
  {\cal H}_j & t_l\sigma_0 & A & \mathbf{0} \\
  t_l\sigma_0 & {\cal H}_j & \mathbf{0} & A \\
  A^\dag & \mathbf{0} & {\cal H}_{4-j} & -t_l\sigma_0 \\
  \mathbf{0} & A^\dag & -t_l\sigma_0 & {\cal H}_{4-j} \\
 \end{pmatrix}
 \right],
\end{equation}
where $A=t_z\sigma_0(1 + e^{-ik_z})$ as before, and $t_l$ is a term coupling layers which host edge modes propagating in the same direction. The mirror symmetry of Eq.~\eqref{eq:reflop} still holds, but now there are two pairs of counter-propagating modes in each unit cell, so we expect a total of four Dirac cones to form on each surface of the slab. For $t_l=0$, the system consists of two independent copies of the AFTCI in Eq.~\eqref{eq:faftci}, producing overlapping surface Dirac cones both at $\varepsilon=0$, $k_x=\pi$ and at $\varepsilon=\pi$, $k_x=0$. Turning on the hopping between co-propagating edge modes, $t_lT=0.3$, moves the Dirac cones away from each other in a symmetric way, since ${\cal F}_2$ obeys the same particle-hole symmetry Eq.~\eqref{eq:AFTCI_phs} as the model with two layers in a unit cell. This is shown in Fig.~\ref{fig:tcipm2}, where we set $JT/3=\pi/2$ and $t_zT=0.6$, and plot in  the bandstructure of the model in an infinite slab geometry consisting of $W=20$ unit cells along the $a_y$ direction.

\begin{figure}[tb]
 \includegraphics[width=0.5\columnwidth]{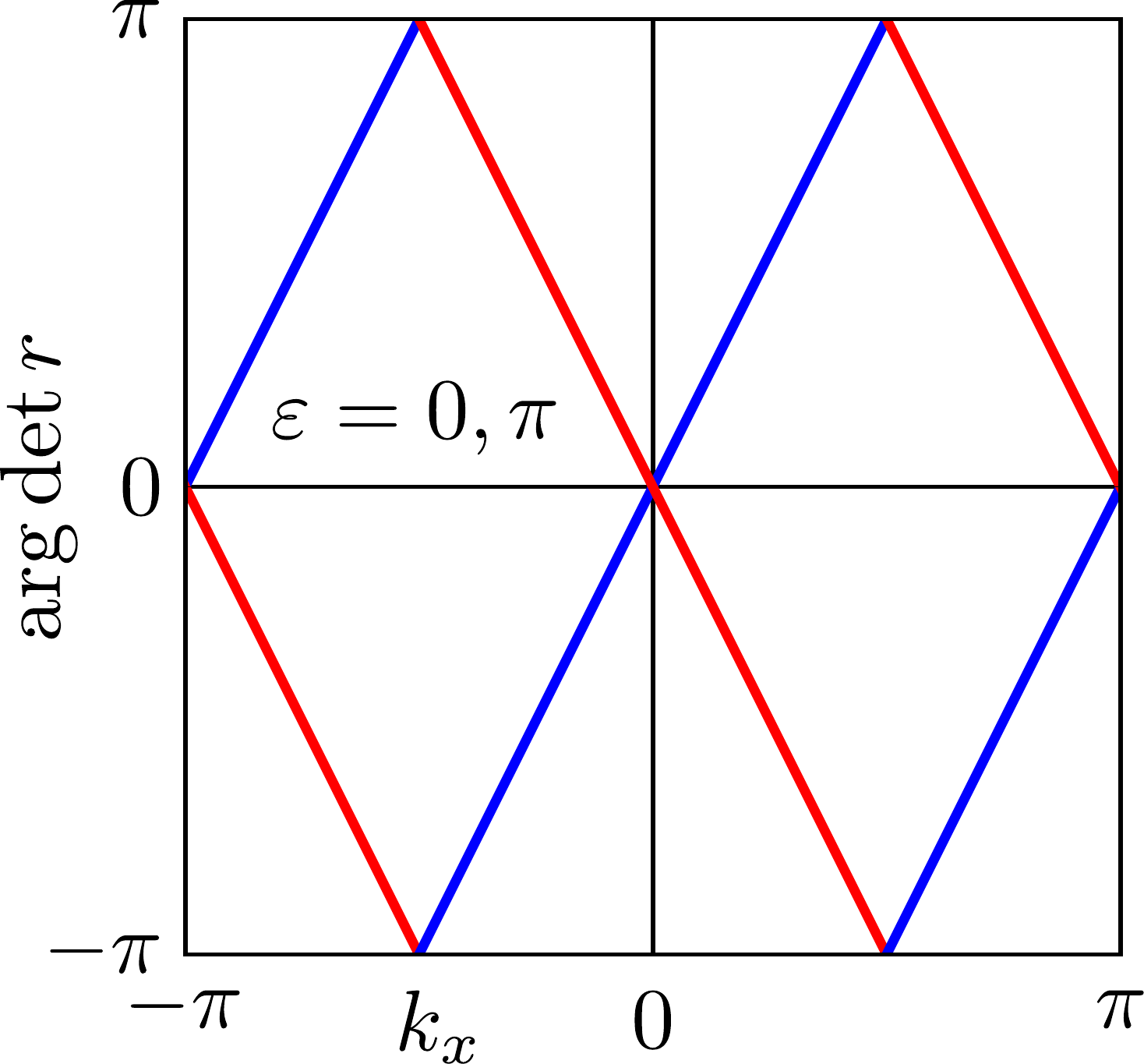}
 \caption{Winding of the phase of $\det r_\pm$ as a function of $k_x$ for the model ${\cal F}_2$, whose bandstructure is shown in Fig.~\ref{fig:tcipm2}. The mirror winding number is ${\cal W}_m=2$ in this case, but the weak mirror indexes vanish both at $k_x=0$ and at $k_x=\pi$.\label{fig:inv_tci_2}}
\end{figure}

The phases of the reflection block determinants are shown in Fig.~\ref{fig:inv_tci_2}, indicating a doubled mirror winding number. However, unlike the previously studied model, here all weak mirror indexes are trivial, $\nu_m(k_x,\varepsilon)=0$, indicating that there are always an even number of surface Dirac cones at high symmetry points in the BZ. This number is two when $t_l=0$, such that two Dirac cones overlap at high symmetry points, and drops to zero as the Dirac cones are shifted away from $k_x=0,\pi$ by a non-zero $t_l$. The above example emphasizes the fact that the values of $\nu_m$ are in general insufficient to describe the exact position of the surface modes, unlike the simpler model of ${\cal F}_{\rm AFTCI}$. This is due to the fact that $\nu_m$ only counts the \textit{parity} of the number of surface Dirac cones at a given high-symmetry point in the bandstructure.

The mirror invariants ${\cal W}_m$ and $\nu_m$ are in fact not independent, but they are related to each other in a similar way to the strong and weak indexes of time-independent topological superconductors\cite{R2010}
\begin{equation}\label{eq:strong_weak}
 (-1)^{{\cal W}_m(\varepsilon)} = (-1)^{\nu_m(0, \varepsilon)+\nu_m(\pi, \varepsilon)},
\end{equation}
which applies both to $\varepsilon=0$ and $\varepsilon=\pi$. The relation of Eq.~\eqref{eq:strong_weak} can be understood as a geometric constraint on the shape of the quasienergy bands in a particle-hole symmetric system. As long as the full set of states must be symmetric under $({\bf k}, \varepsilon)\to(-{\bf k}, -\varepsilon)$, then for any odd number of boundary modes, at least one should be pinned to a particle-hole symmetric point of the BZ, as happens in Figs.~\ref{fig:hex} and \ref{fig:tci1}. However, if the total number of boundary modes is even, it is possible that none of them exist at high symmetry points, as shown in Fig.~\ref{fig:tcipm2}. Note that this does not exhaust all possible options for Dirac cone positions. It is possible for a system to host two Dirac cones, each of which sits at particle-hole symmetric position, in which case both of the weak mirror invariants take non-trivial values.

\begin{figure}[tb]
 \includegraphics[width=\columnwidth]{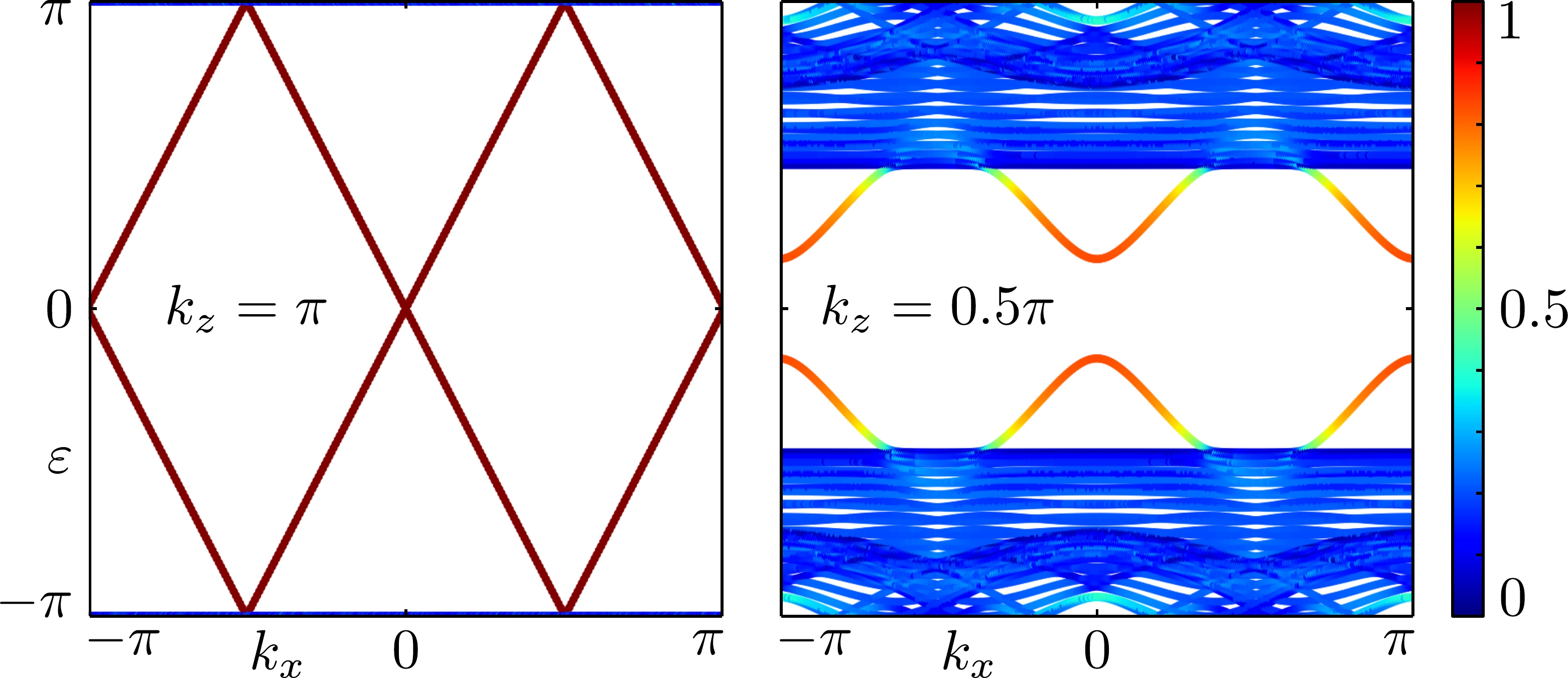}
 \caption{Folded bandstructure of ${\cal F}_{\rm AFTCI}$ in the quasienergy direction, obtained by doubling the number of steps in the driving protocol, as described in the main text. The system geometry and parameter values are the same as Fig.~\ref{fig:tci1}. The two original dispersionless bulk bands now overlap at the new BZ boundary, $\varepsilon=\pi$, and there are two  Dirac cones on each surface, in the gap at $\varepsilon=0$. Unlike Fig.~\ref{fig:tcipm2}, each of the two cones is separately pinned to a high-symmetry point of the BZ.\label{fig:tcipm1_doubled}}
\end{figure}

\begin{figure}[tb]
 \includegraphics[width=0.5\columnwidth]{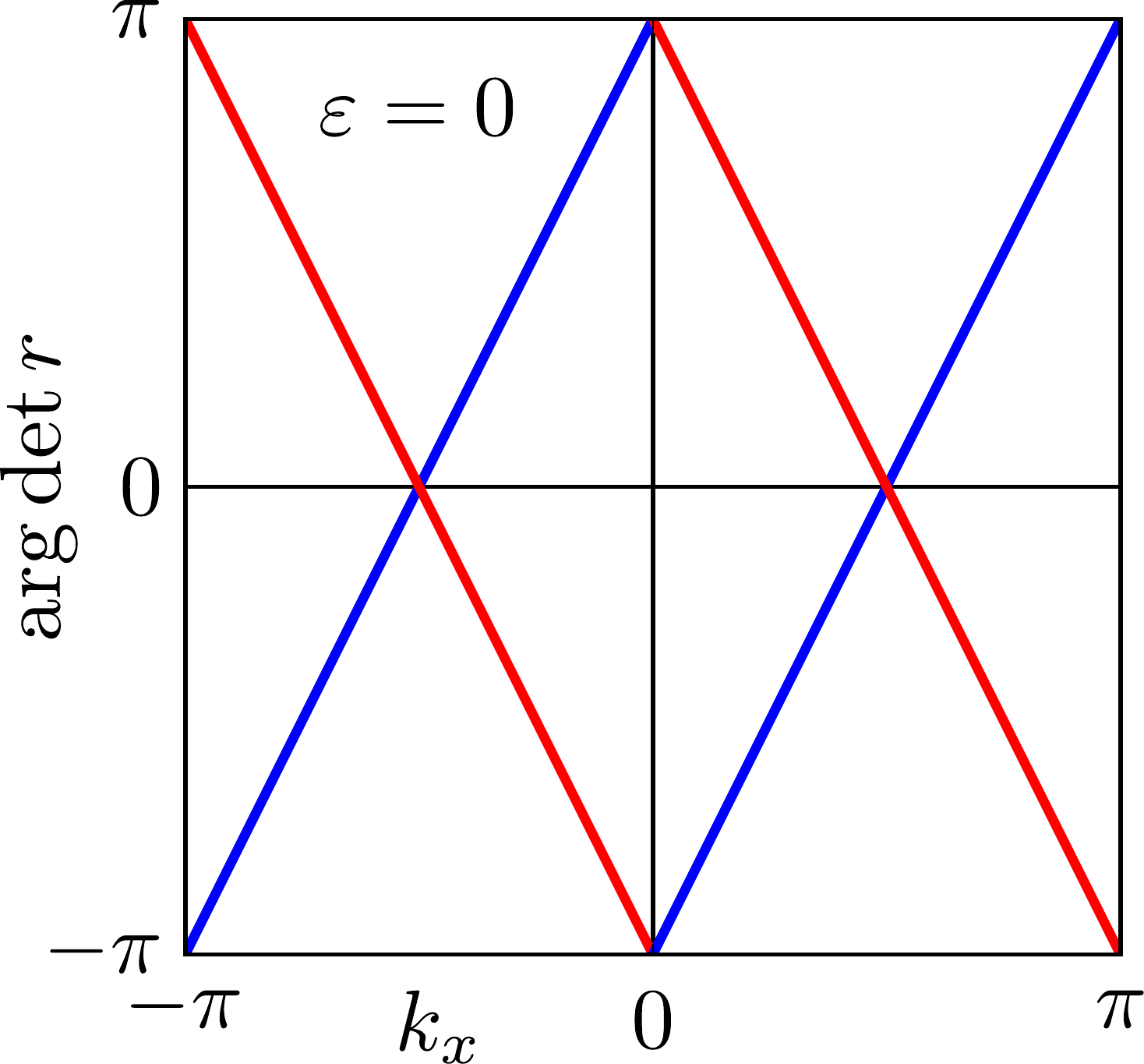}
 \caption{Winding of the phase of $\det r_\pm$ for the time folded model with the bandstructure of Fig.~\ref{fig:tcipm1_doubled}, at quasienergy $\varepsilon=0$. The color code is as before, blue and red for the phases of the determinants of $r_+$ and $r_-$, respectively. We find a value ${\cal W}_m=2$, but now both mirror weak indexes are non-trivial $\nu_m(0,0)=\nu_m(\pi,0)=1$, consistent with the separate pinning of each Dirac cone to a high symmetry point.\label{fig:inv_tci_1_doubled}}
\end{figure}

To show an example of such a phase, we take the simple model ${\cal F}_{\rm AFTCI}$ of Eq.~\eqref{eq:faftci} and increase the number of surface modes not by making the unit cell bigger, but by doubling the number of steps in the driving protocol. This procedure was referred to as \textit{time folding} in Ref.~\onlinecite{Fulga2016a}, since doubling the number of steps and the period of the time evolution leads to a folding of the BZ in quasienergy, similar to how doubling the unit cell in real space folds the BZ in momentum. We consider a new driving cycle consisting of 6 steps, where in each half cycle the hoppings $J_x$, $J_y$, and $J_z$ are turned on successively, as described in Section \ref{sec:aftci}. As before, to ensure that neighboring layers have chiral modes propagating in opposite directions, we invert the sequence of steps for the $H_-$ layers as compared to the $H_+$ ones. The resulting bandstructure is shown in Fig.~\ref{fig:tcipm1_doubled} and the phases of the reflection block determinants in Fig.~\ref{fig:inv_tci_1_doubled}. The two flat bulk bands of the original model are now folded on top of each other at the quasienergy zone boundary of the new BZ, $\varepsilon=\pi$. Similarly, the two surface Dirac cones now appear in the same gap, $\varepsilon=0$. However, unlike the case of Fig.~\ref{fig:tcipm2}, each of the Dirac cones is separately pinned to a high symmetry point of the BZ due to particle hole symmetry. This constraint is evidenced by the topological invariants of the folded system, which read ${\cal W}_m=2$ and $\nu_m(0,0)=\nu_m(\pi,0)=1$. In principle, successively doubling the driving period allows for the generation of an arbitrarily large number of topological boundary modes, similar to the case of 2D Floquet topological phases reported in Refs.~\onlinecite{Ho2014, Zhou2018a}.

\section{Conclusion}
\label{sec:conc}

We have extended the notion of a 3D topological crystalline insulator to the periodically driven setting, by introducing a new class of Floquet mirror symmetry protected systems. Due to their time dependent nature, we have shown that anomalous topological phases can be realized, in which the relevant TCI index, the mirror Chern number, does not capture the topological protection of surface Dirac cones. Instead, we have adapted the scattering matrix formulation of topological invariants to TCIs, and introduced a mirror winding number which correctly expressed the bulk-boundary correspondence of these Floquet models. Since the 2D winding number Eq.~\eqref{eq:s_inv} was shown to provide a unified description of both static and driven systems,\cite{Fulga2016a} we expect that the mirror index Eq.~\eqref{eq:mirror_W} should also be valid in describing time-independent TCIs, taking values equal to the mirror Chern number.
Further, we have shown that in the presence of particle-hole symmetry, a new topological invariant containing information on the position of boundary modes can be introduced. This mirror weak index [Eq.~\eqref{eq:mirror_nu}] is obtained in a similar way to the mirror Chern number, and determines the parity of the number of modes at particle-hole symmetric points in the surface BZ.

In time-independent systems, the seminal works which introduced the concept of a TI protected by point-group symmetries\cite{Teo2008, Fu2011, Hsieh2012} almost a decade ago have led to an intense and diverse research pursuit in classifying TCIs and analyzing their properties.\cite{Tanaka2012, Dziawa2012, Xu2012, Slager2012, Chiu2013, Morimoto2013, Jadaun2013, Zhang2013, Kargarian2013, Benalcazar2014, Wrasse2014, Fang2015, Ando2015, Diez2015, Morimoto2015, Kim2015, Zhou2018, Song2018} In contrast, the effect of point-group symmetries has yet to be thoroughly investigated in the context of time-periodic systems. In this regard, Floquet TIs protected by rotation symmetry as well as the topological classification of Floquet TCIs are interesting topics for future work. By showing how Floquet TCIs can be readily constructed from well-understood, lower-dimensional building blocks, we hope that our work will motivate further research in these directions.

One of the advantages of constructing 3D models out of stacks of topologically non-trivial layers is that in the resulting anomalous Floquet TCIs both the number and the position of surface Dirac cones can be tuned. Similar to the static case, this can be achieved by changing the number of layers in a unit cell and by modifying the inter-layer coupling terms. However, periodic driving also provides more routes to manipulate the surface modes, which are not possible in a time-independent system. As we have shown, doubling the number of steps in the driving period leads to surface BZ which is folded in quasienergy, providing a means to double the number of Dirac cones while preserving the anomalous nature of the system. In some cases, this procedure could be applied successively, each time increasing the mirror winding number.

Finally, it is interesting to consider how scattering matrix invariants may help to describe the bulk-boundary correspondence of other types of TCI, both static and driven. As shown in Ref.~\onlinecite{Trifunovic2017}, scattering matrices can be used to derive the topological classification of mirror symmetry protected TCIs, and a similar approach should be valid also for other symmetries, such as rotation or glide symmetry. The same idea of block-diagonalizing the reflection matrix, as done in Eq.~\eqref{eq:rblock_diag}, would then lead to new invariants which determine the topological boundary states of Floquet and time-independent systems.

\acknowledgements

We thank Ulrike Nitzsche for technical assistance.

\bibliography{AFTCI_bib}

\newpage
\appendix

\section{Distinguishing between overlapping Dirac cones}
\label{app:DCpos}

\begin{figure}[h]
 \includegraphics[width=0.75\columnwidth]{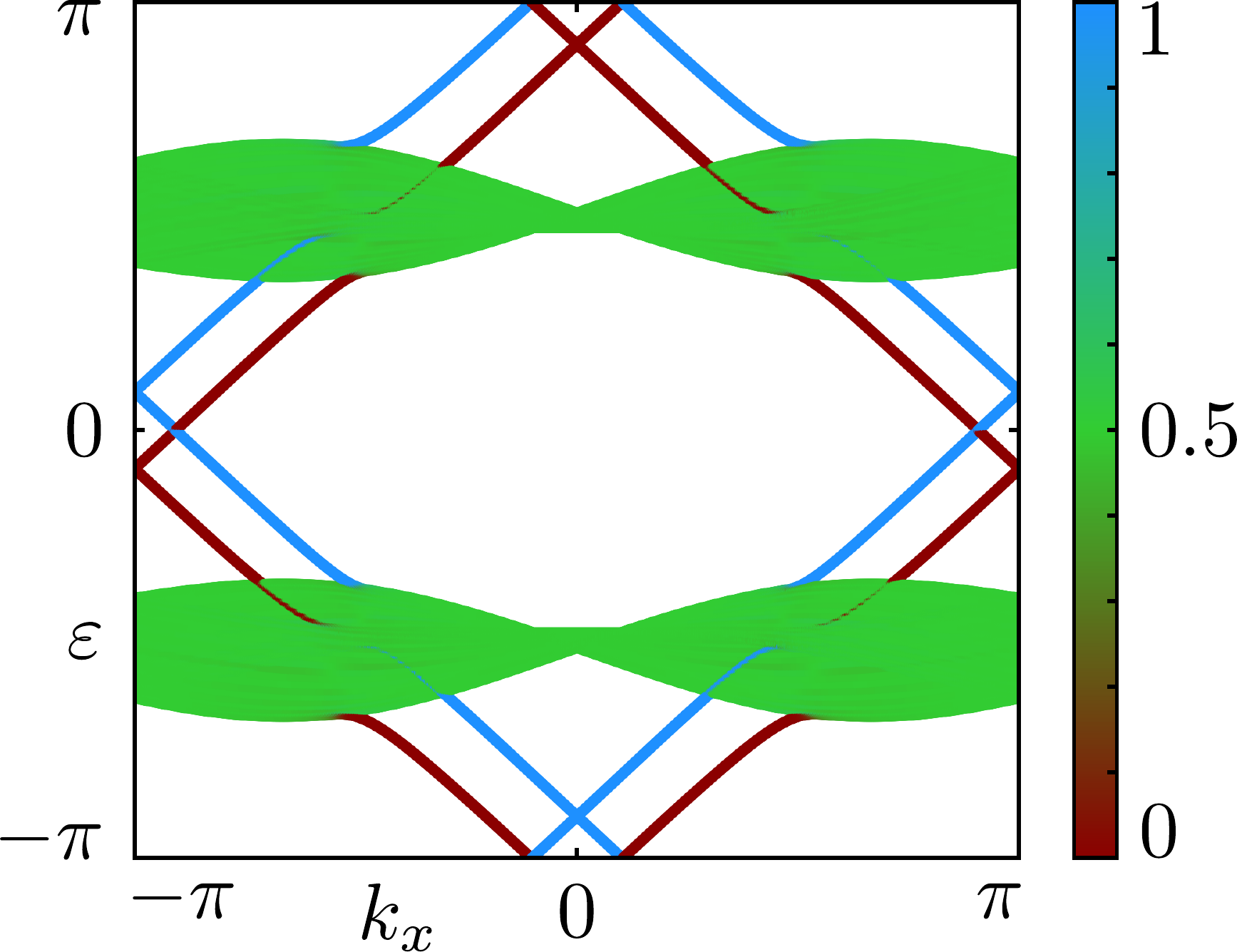}
 \caption{Effect of a staggered chemical potential term $\mu$ on the bandstructure of the anomalous Floquet TCI. We use $\mu T=0.9$, $k_z=\pi$, whereas all other parameters and the same slab geometry are the same as in Fig.~\ref{fig:tci1}. The color scale represents the probability amplitude of states integrated over half of the slab unit cells, corresponding to $1\leq n_y\leq W/2$. As such, states localized on opposite surfaces are shown in blue and red, whereas bulk states are shown in green. There are a total of two Dirac cones on each surface, which are now shifted away from $\varepsilon=0,\pi$ as a consequence of the fact that the chemical potential term breaks particle-hole symmetry. \label{fig:DCshift}}
\end{figure}

The bandstructure of Fig.~\ref{fig:tci1} is computed in a slab geometry consisting of 20 unit cells in the $a_y$ direction, and as such the system contains two surfaces. Each of them hosts a total of two surface Dirac cones, one at $\varepsilon=0$ and another at $\varepsilon=\pi$, which overlap in both quasienergy and momentum. To distinguish between the modes appearing on each surface, here we add a staggered chemical potential term to the layered system. Specifically, we include a positive, time-independent chemical potential term $+\mu\sigma_0$ on all of the even layers of the AFTCI and the opposite term $-\mu\sigma_0$ on all of the odd layers ($\mu T=0.9$). While this term breaks particle-hole symmetry, it enables to better visualize the number of Dirac cones present on each surface, since they are shifted in quasienergy, as shown in Fig.~\ref{fig:DCshift}.

\end{document}